# Revealing the Charge Transfer Dynamics Between Singlet Fission Molecule and Hybrid Perovskite Nanocrystals

Tejasvini Sharma,[1] Saurav Saini,[2] Naveen Kumar Tailor,[1] Mahesh Kumar,[2] Soumitra Satapathi,[1,*]

[1] Department of Physics, Indian Institute of Technology Roorkee, Roorkee, Haridwar, 247667, India
[2] National Physical Laboratory, New Delhi, India

**ABSTRACT:** Singlet fission process has gained considerable attention because of its potential to enhance photovoltaic efficiency and break the Shockley–Queisser limit. In photovoltaic devices, perovskite materials have shown tremendous progress in the last decade. Therefore, combining the singlet fission materials in perovskite devices can lead to a drastic enhancement in their performance. To reveal the applicability of singlet fission processes in perovskite materials, we have investigated the charge transfer dynamics from an SF active material 9,10-bis(phenylethynyl)anthracene to $CH_3NH_3PbBr_3$ perovskite nanocrystals using the transient absorption spectroscopy. We observed a significant charge transfer from the coupled $^1TT$ state of BPEA to conduction band of $CH_3NH_3PbBr_3$ in picosecond timescale. The observation of shortened lifetime in a mixture of BPEA and $CH_3NH_3PbBr_3$ nanocrytals confirms the significant charge transfer between these systems. Our study reveals the charge transfer mechanism in singlet fission-perovskite composite which will help to develop an advanced photovoltaic system.

The singlet fission (SF) process has been gained considerable attention in recent years because it promises to break the conventional Shockley–Queisser limit.[1-2] In the singlet fission process, two triplet excitons are generated from one singlet exciton.[3-4] The SF process is widely reported in the various type of organic molecules such as TIPS-pentacene, anthracene, and its derivatives like 9,10-bis(phenylethynyl)anthracene (BPEA), tetracene, etc.[5-11] Despite significant advances in understanding the chemical mechanism that drives the SF process, practical application in photovoltaic devices remains a challenge. Further development in this field requires the integration of SF active molecules with chromophores which will be able to extract the photoexcited carriers from SF molecules leading to enhanced device performance. Several attempts have been attempted to fabricate SF-molecular chromophores dyads and to study charge/energy transfer dynamics at the interface. In this manner, Diaber et.al. demonstrated the triplet transfer from tetracene to Silicon with a charge transfer efficiency of 36%.[12] Additionally, MacQueen et.al. fabricated silicon heterojunction solar cells with SF material, tetracene.[13]

Recently, organic-inorganic metal halide perovskites have attracted immense interest in photovoltaics due to wide absorption region, tunable band gap, efficient charge generation, high luminescence yields, high-performance, and low-cost solution-based fabrication processes.[14-22] It has been shown that effective triplet energy transfer at organic–inorganic NC interfaces sensitises the creation of long-lived triplet states that may be used for photon upconversion, photocatalysis, and increased light emission. Ehrler et. al calculated that when SF molecules are coupled with inorganic quantum dots, that photons may be transferred into the underlying solar cell, boosting the limiting PCE by 4.2% absolute.[23] The molecules absorb light-producing singlet states, which are subsequently converted to triplet states by SF, and the triplets are then transferred to the inorganic NCs. Furthermore, Lee et al. also investigated carrier dynamics in a TIPS-Pentacene/MAPbI$_3$ bilayer film and hypothesised that electron transfer occurs from TIPS-PEN triplet states to the MAPbI3 perovskite.[24] Later, Guo et al. demonstrated the transfer of electron within 1.5 ps from the correlated bound triplet pair states of TIPS-pentacene to conduction band of MAPbI$_3$ perovskite films using transient absorption spectroscopy.[25] However, exploration of the new perovskite chromophores and study of charge transfer between singlet fission molecule and perovskite system remains limited in the literature and needs more in-depth investigation.

Here, we have investigated electron/hole-transfer dynamics between BPEA, a well-known singlet fission material, and CH$_3$NH$_3$PbBr$_3$ (MAPbBr$_3$) perovskite nanocrystals as an electron acceptor material to reveal their suitability for efficient charge/energy transfer. We have chosen the MAPbBr$_3$ nanocrystal because of its visible fluorescence and its recent use in solar cells. The 9,10-bis(phenylethynyl)anthracene (BPEA) as an SF active molecule has been actively used in light-emitting diodes and solar cells. The singlet and triplet energy levels in BPEA appear to be perturbed by phenylethynyl substituents, making the SF process thermodynamically stable. The MAPbBr$_3$ perovskite nanocrystals were synthesized using the ligand-assisted reprecipitation (LARP) technique in the chlorobenzene antisolvent and BPEA solution was also prepared in the chlorobenzene. Optical properties were investigated using UV vis absorption spectroscopy and photoluminescence spectroscopy. The excited-state dynamics and charge transfer dynamics were probed using the femtosecond transient absorption (TA) spectroscopy on pristine $MAPbBr_3$, BPEA, and the mixture of BPEA and $MAPbBr_3$ nanocrystals. We observed charge transfer from coupled $^1TT$ state of BPEA to the conduction band of the $MAPbBr_3$ nanocrystal system in a picosecond timescale. Our study provides insights into the charge/energy transfer mechanism between BPEA and perovskite NCs which will help to develop highly efficient perovskite-based devices utilizing the SF molecules.

The $MAPbBr_3$ perovskite nanocrystals were synthesized using a LARP technique (Figure S1a).[26-27] Precipitation occurs in this approach due to reduced solubility caused by antisolvent mixing in the presence of capping ligands on the surface, which restricts further particle growth and aggregation. The images of obtained nanocrystals is shown in the Figure S1b. Under UV irradiation, the solution progressively changed colour from transparent to bluish-green, indicating the development of nanocrystals that produce a strong blue-green light (inset of Figure S1b). The chlorobenzene was used as an antisolvent for nanocrystal fabrication. For preventing any solution effect, we used the same chlorobenzene for preparing the BPEA solution. The BPEA solution is highly fluorescent and gives the emission in the visible region. The crystal structure of $MAPbBr_3$ perovskite and BPEA is presented in Figures 1a and 1b, respectively. The optical properties of as-synthesized $MAPbBr_3$ NCs and BPEA are investigated by steady-state UV-Vis absorption spectroscopy and photoluminescence (PL) spectroscopy and are presented in Figure 1(c-e). The absorption spectra of $MAPbBr_3$ NCS (blue curve in Figure 1c) exhibit a band-edge around 515 nm which is consistent with the previous reports. The emission spectrum (red curve in Figure 1c) shows the peak around 520 nm (2.38 eV) with the FWHM of 30 nm when excited at 350 nm. Further, the absorption spectrum of BPEA in chlorobenzene exhibits three peaks at 462, 440, and 312 nm (Figure 1d).[28-29] The absorption peak centered at 312 nm due to $π–π^*$ transition is ascribed to the short axis of the anthracene moiety. The peak centered at 440 nm are due to the transition of the long axis of anthracene.[28-29] The emission spectrum of BPEA in chlorobenzene exhibits peaks at 483 and 510 nm, which originates from the BPEA monomer. We observed that photoluminescence spectra are not exactly the mirror image of absorption spectra in BPEA solution, which can be attributed to the existence of a variety of conformations of BPEA in which the two phenyl groups can rotate almost freely in the solution at room temperature. Interestingly, while we mixed the BPEA solution into perovskite NCs solution (1: 50 volume ratio), the absorption of BPEA dominates because of its high absorption coefficient (Figure 1e). In the emission spectra, the second peak of BPEA (510 nm) dominates and an extra shoulder appears around 550 nm.

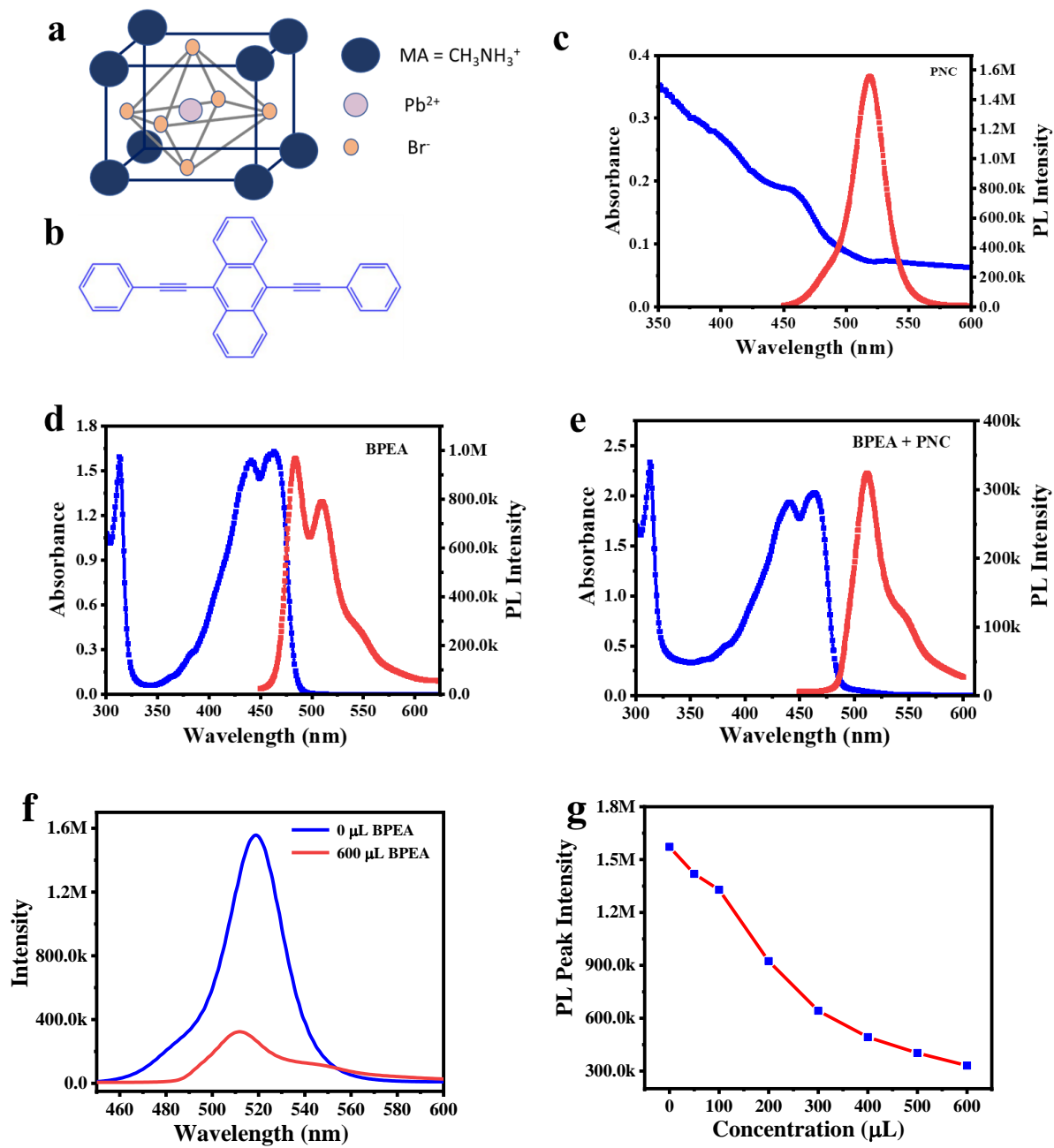

**Figure 1.** Schematic crystal structure of (a) MAPbBr$_3$ and (b) 9,10-Bis(phenylethynyl)anthracene (BPEA). Steady-state absorbance and photoluminescence spectra of (c) MAPbBr$_3$ nanocrystals dispersed in chlorobenzene solution, (d) BPEA solution in chlorobenzene, and (e) mixture of BPEA and MAPbBr$_3$ nanocrystals. The blue curve represents the absorbance and the red curve represents the emission spectrum. (d) The reduction in photoluminescence of NCs with the addition of BPEA. (f) The PL peak intensity with the increasing concentration of BPEA in NCs solution.

Furthermore, to observe the effect on emission characteristics of PNCs after adding the BPEA, we gradually mixed the BPEA solution into the NCs solution (0 μL to 600 μl). We observed that the emission intensity of PNCs significantly reduces with the addition of BPEA as shown in Figure 1f. With the 600 μL addition of BPEA, the emission intensity was quenched more than 5 times. The intensity reduction with adding of BPEA has started from 0 μL to 600 μL as shown in Figure 1g. The significant reduction in the PL intensity of NCs with BPEA addition suggests several possible photoinduced processes including charge or energy transfer.[30-31]

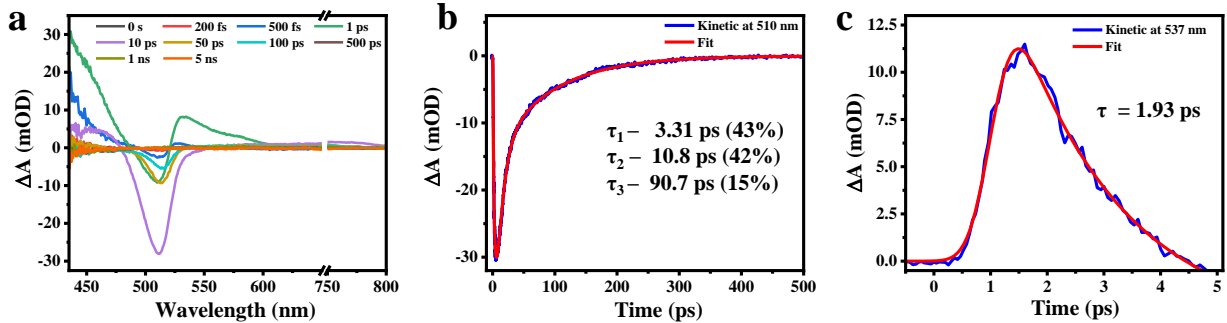

**Figure 2.** (a) The femtosecond transient absorption spectrum of MAPbBr$_3$ NCs excited at 350 nm wavelength. (b, c) fitted kinetics at (b) 509 nm and (c) 537 nm. The red line represented the fitted data. The lifetime components with their percentage weightage are given in the inset of the figures.

To investigate the details of charge or energy transfer dynamics, we have measured femtosecond transient absorption (fs-TA) spectra on MAPbBr$_3$ NCs, BPEA solution, and the chromophore mixture of MAPbBr$_3$ NCs and BPEA. In all these cases, the pump wavelength was at 350 nm and the probe wavelength was used in the visible wavelength (400 - 800 nm) region. The transient absorption spectra of MAPbBr$_3$ nanocrystals are shown in Figure 2a. The corresponding contour plot is shown in Figure S2a. We observed the ground state bleach (GSB) signal in the range of 470 to 530 nm, which directly relates to the depletion of ground-state charge carriers due to band filling.[32] It is interesting to observe that the ground-state bleach (GSB) peak red shifts with increasing the delay time upto 5 ps. Similarly, the intensity and width of the GSB signal are also enhanced with an increased delay of up to 5 ps. The broadening and position shift of the GSB signal can be attributed to the coulombic screening and bandgap renormalization effects.[33-34] The kinetics for GSB at 509 nm is shown in Figure 2b and lifetime components were obtained as $\tau_1$ = 3 ps (41%), $\tau_2$ = 11.9 ps (42%) and $\tau_3$ = 92.4 ps (17%). The first component $\tau_1$ = 3 ps can be attributed to the hot carrier relaxation.[33, 35] The second component $\tau_2$ = 11.9 ps can be attributed to the charge carrier trapping to the sub-bandgap trap state and the third component $\tau_2$ = 92.4 ps can be assigned to the exciton recombination.[36-37] Apart from GSB, the TA spectra of MAPbBr$_3$ NCs also contain a positive signal centered around 535 nm which can be attributed to the photo-induced emission. The kinetics for the PIA band at 537 nm is presented in Figure 2c and the lifetime component was obtained as $\tau$ = 1.93 ps which can be assigned to the ultrafast hot carrier relaxation.[32, 35] It is noteworthy to observe that the PIA band decreases very fast and converts into a bleach signal and this bleach signal recovers in 30 ps as presented in Figure S2b.

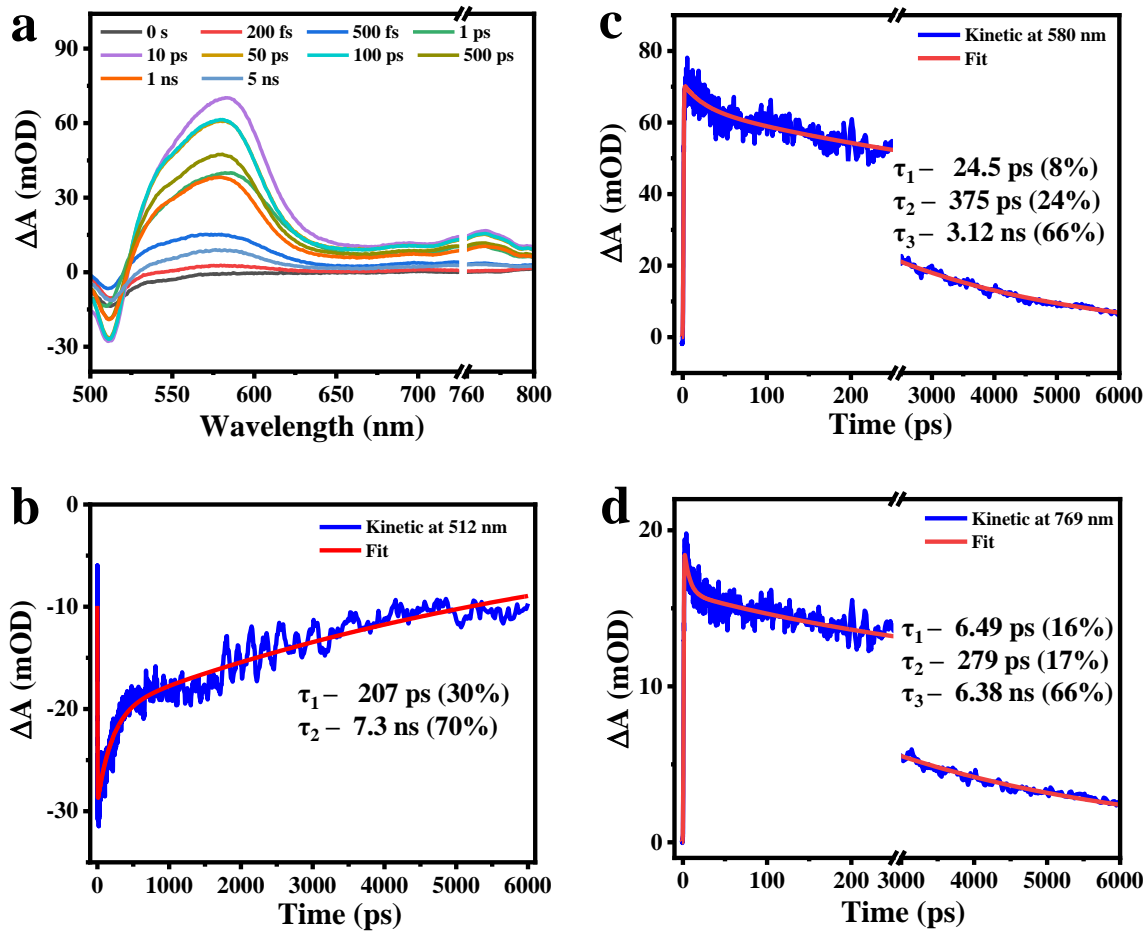

**Figure 3.** (a) The femtosecond transient absorption spectrum of BPEA solution excited at 350 nm wavelength. (b-d) Fitted kinetics at (b) 512 nm, (c) 580 nm and (d) 769 nm. The red line represented the fitted data. The lifetime components with their percentage weightage are given in the inset of the figures.

Further, we have recorded the TA spectra of the BPEA solution as presented in Figure 3a and corresponding contour plot is shown in Figure S3. The TA spectra shows a negative band around 500−525 nm, which corresponds to ground state bleach and a positive excited state absorption (ESA) band associated with the S1 → Sn transition in the 525 - 650 nm region centered at 580 nm.[38-39] The TA spectra show a negative band at 500 - 525 nm owing to a ground state bleach, and a positive excited-state absorption (ESA) band in the 525 - 650 nm area centered at 580 nm related to the S1 Sn transition. We found that the GSB signal builds up in 100 ps and the ESA signal builds up in 10 ps. The ESA band shows a modest redshift with increasing intensity up to 10 picoseconds, which can be attributed to vibrational cooling and planarization of the phenylethynyl rings in the S1 state. After 10 picoseconds, the intensity reduces on several nanosecond time scales. The kinetics for GSB at 512 nm is presented in Figure 3b and lifetime components were obtained as $\tau_1$ = 207 ps (30%), and $\tau_2$ = 7.3 ns (70%). The kinetics for ESA band at 580 nm is presented in Figure

3c and $\tau_1 = 24.5$ ps (8%), $\tau_2 = 375$ ps (24%) and $\tau_3 = 3.12$ ns (66%). Additionally, we also observed a small positive signal centered at 769 nm, which also decays in nanosecond timescale (Figure 3d).

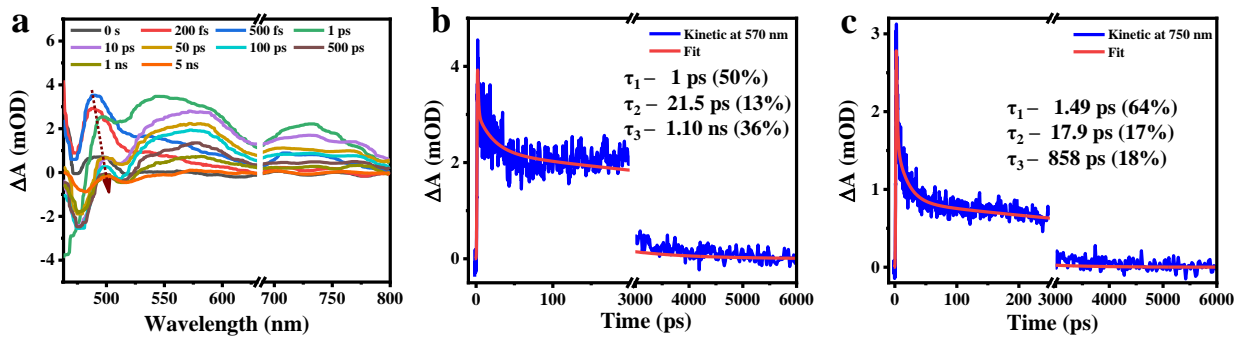

**Figure 4.** (a) The femtosecond transient absorption spectrum of the mixture of NCs and BPEA solution excited at 350 nm wavelength. (b,c) Fitted kinetics at (b) 570 nm and (c) 750 nm. The red line represented the fitted data. The lifetime components with their percentage weightage are given in the inset of the figures.

After studying individual excited states of the chromophores, we proceed to measure the charge/energy transfer dynamics of their mixture as this will be useful for developing photovoltaics devices. We have observed a significant change in the dynamics of the PNC and BPEA mixture when excited at 350 nm, containing multiple positive and negative signals because of the combined effect of BPEA and PNCs (Figure 4a and S4). The analysis of the mixture of BPEA and PNCs TA spectra reveals the possibility of charge transfer from BPEA to MAPbBr$_3$ NCs system (Figure 4b and 4c).[25] Particularly, we have focused our discussion on ESA bands at 570 and 750 nm and analyzed the kinetics. The obtained lifetime components are presented in Table 1.

**Table 1.** Lifetime component for the kinetics of MAPbBr$_3$ NCs, BPEA, and mixture of MAPBbr$_3$ and BPEA.

| Sample | Wavelength (nm) | $\tau_1$ | $\tau_2$ | $\tau_3$ | Avg. Lifetime |
|---|---|---|---|---|---|
| MAPbBr$_3$ | 509 nm | 3.31 ps | 10.8 ps | 90.7 ps | 19.56 ps |
|  | 537 nm | 1.93 ps | - | - | 1.93 ps |
| BPEA | 512 nm | 207 ps | 7.3 ns | - | 5.17 ns |
|  | 580 nm | 24.5 ps | 375 ps | 3.12 ns | 2.17 ns |
|  | 769 nm | 6.49 ps | 279 ps | 6.38 ns | 2.44 ns |
| MAPbBr$_3$ + BPEA | 570 nm | 1 ps | 21.5 ps | 1.10 ns | 402 ps |
|  | 750 nm | 1.49 ps | 17.9 ps | 858 ps | 161 ps |

It is well known that for efficient charge transfer, energy level matching is a main criterion.[25] The highest occupied molecular orbital (HOMO) energy of the BPEA is -5.49 eV and the lower bound for the lowest unoccupied molecular orbital (LUMO) energy of -2.92 eV as taken from the previous literature.[28, 38-39] The energy level of valence and conduction band of MAPbBr$_3$ were taken as -5.8 and -3.4 eV, respectively. The energy level diagram of the BPEA/MAPbBr$_3$ system is depicted in Figure 5 which clearly shows the possibility of single-singlet transition. Furthermore, for the charge transfer from the triplet state of BPEA to perovskite NC, the energy level of the triplet state of BPEA should be higher than the conduction band (CB) energy of MAPbBr$_3$ NCs. Previously, it is reported that the triplet state energy of BPEA is E(T1) = 1.1 eV. This energy level is below the CB energy level of the MAPbBr$_3$ system. As a result, we may deduce that charge transfer from triplet states of BPEA to conduction band of MAPbBr$_3$ will be thermodynamically unfavourable. Electron transfer from the correlated bound triplet pair state (denoted by 1(TT)) formed by SF in BPEA to the MAPbBr$_3$ conduction band is another proposed charge transfer mechanism. In materials with high SF efficiencies, it has previously been observed that the 1(TT) state is essentially iso-energetic with S1 or double the energy of T1. Therefore, we calculate the upper limit of energy of 1(TT) state to be 2.2 eV of the BPEA. Therefore, it is expected that electron transfer from coupled triplet state to the conduction band of perovskite is prevailing in this case.

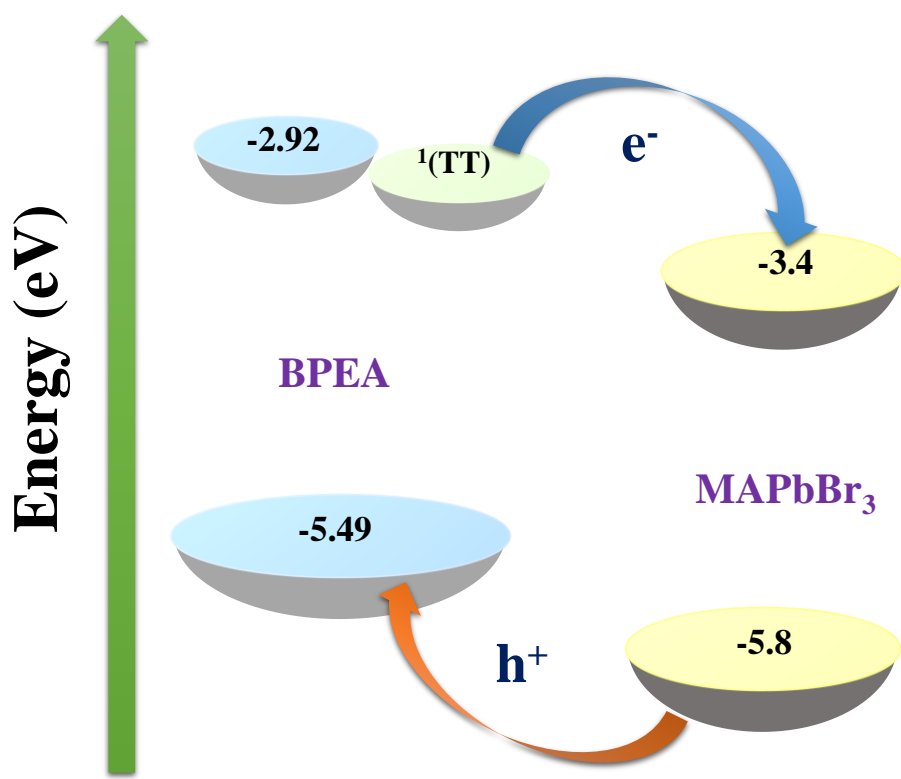

**Figure 5.** Schematic diagram of the charge-transfer dynamics between BPEA and MAPbBr$_3$ perovskite nanocrystals.

The above analysis is further confirmed by comparing the TA kinetics at $\lambda_{probe}$ = 570 in a mixture of BPEA and MAPbBr$_3$ NCs. The band around 580 nm of BPEA was attributed to coupled triplet state 1(TT) states of the BPEA, according to its energy level (2.2 eV).[40] After mixing the BPEA in PNCs, the band is blue-shifted to 570 nm. The fast lifetime component $\tau_1$ = 24.5 ps of BPEA is reduced to $\tau_1$ = 1 ps in the BPEA-MAPbBr$_3$ mixture. In addition, if we compare the average lifetime of the 580 nm band for BPEA and BPEA and NC mixture, the lifetime has reduced from 2.17 ns to 402 ps. As a result, we infer that ultrafast electron transfer occurs in picosecnd timescale from the correlated triplet pair state generated via SF in BPEA to the conduction band of MAPbBr$_3$.[24-25, 38-39, 41] It's also possible that because of the spectral overlap between 1(TT) and T1 in the visible range. it is usually difficult to entirely separate the spectral characteristics of T1 and 1(TT). Additionally, we found that the band at 769 nm in the BPEA solution is blue-shifted to 750 nm in the case of the BPEA-PNCs mixed solution and the average lifetime is also reduced from 2.44 ns to 161 ps, which also suggests the significant charge transfer from BPEA to PNCs. It is noteworthy to observe that the charge transfer route from coupled triplet state is substantially less efficient than the direct triplet pair charge transfer route.

In summary, we have demonstrated the charge transfer dynamics in an SF active material BPEA and perovskite MAPbBr$_3$ nanocrystals (NCs) using ultrafast femtosecond transient absorption spectroscopy. The MAPbbr3 NCs were prepared using the LARP technique. Both the BPEA and NCs give the emission in the visible range. We found a significant reduction in the fluorescence intensity in the mixture of BPEA and NCs which indicates the charge transfer between these systems. The charge transfer was further verified with the transient absorption measurement. We observed a significant charge transfer from coupled $^1$TT state of BPEA to the conduction band of the MAPbBr$_3$ NCs system in a picosecond timescale. The observation of shortened lifetime in a mixture of BPEA and NCs confirms the charge transfer mechanism between these systems. The understanding of the charge transfer dynamics in SF material and perovskite systems can help to design perovskite photovoltaic devices based on the SF process and the fabrication of new types of perovskite solar cells in the future.

- **ASSOCIATED CONTENT**
  **Supporting Information**

The Supporting Information is available
Experimental methods, synthesis scheme and TEM image of MAPbBr$_3$ NCs, contour plots of MAPbBr$_3$ NCs, BPEA, and mixture of MAPbBr$_3$ NCs and BPEA

- **ACKNOWLEDGMENTS**

TS like to acknowledge CSIR fellowship.